\documentclass[structabstract]{aa}  
\newcommand{\GILDAS}{\texttt{GILDAS}}
\usepackage{natbib}
\usepackage{amsmath}
\usepackage{graphicx}
\usepackage{txfonts}
\usepackage{subfig}
\usepackage{multirow} 
\usepackage{graphicx}
\usepackage{threeparttable}
\newcommand{\ie}{\emph{i.e.}}
\newcommand{\eg}{e.g.}
\newcommand{\emm}[1]{\ensuremath{#1}}   
\newcommand{\emr}[1]{\emm{\mathrm{#1}}} 
\newcommand{\Tkin}{\emm{T_\emr{kin}}}
\newcommand{\Tex}{\emm{T_\emr{ex}}}
\newcommand{\Trot}{\emm{T_\emr{rot}}}
\newcommand{\Td}{\emm{T_\emr{dust}}}
\newcommand{\nH}{\emm{n_\emr{H}}}
\newcommand{\DCOp}{\emr{DCO^{+}}}                  
\newcommand{\h}{\text{H}}                          
\newcommand{\x}{\text{X}}                          
\newcommand{\hhco}{\text{H}_2\text{CO}}            
\newcommand{\hhhco}{\text{H}_3\text{CO}}            
\newcommand{\chhhoh}{\text{CH}_3\text{OH}}         
\newcommand{\ohhco}{\text{o-H}_2\text{CO}}         
\newcommand{\phhco}{\text{p-H}_2\text{CO}}         
\newcommand{\hhthico}{\text{H}_2^{13}\text{CO}}    
\newcommand{\ohhthico}{\text{o-H}_2^{13}\text{CO}} 
\newcommand{\phhthico}{\text{p-H}_2^{13}\text{CO}} 
\newcommand{\ddco}{\text{D}_2\text{CO}}            
\newcommand{\oddco}{\text{o-D}_2\text{CO}}         
\newcommand{\pddco}{\text{p-D}_2\text{CO}}         
\newcommand{\hh}{\text{H}_2}                       
\newcommand{\ohh}{\text{o-H}_2}                    
\newcommand{\phh}{\text{p-H}_2}                    
\newcommand{\hho}{\text{H}_2\text{O}}              
\newcommand{\unit}[1]{\emm{\, \emr{#1}}}
\newcommand{\mm}{\unit{mm}}
\renewcommand{\deg}{\emm{^\circ}}
\newcommand{\pccm}{~\rm{cm}^{-3}}
\newcommand{\pscm}{~\rm{cm}^{-2}}
\newcommand{\ps}{~\rm{s}^{-1}}
\newcommand{\kms}{~\rm{km}~\mathrm{s}^{-1}}
\newcommand{\Kkms}{~\rm{K\,km\,s}^{-1}}
\newcommand{\Tsys}{\emm{T_\emr{sys}}}
\newcommand{\Tas}{\emm{T_\emr{A}^*}}
\newcommand{\Tmb}{\emm{T_\emr{mb}}}
\newcommand{\Beff}{\emm{B_\emr{eff}}}
\newcommand{\Feff}{\emm{F_\emr{eff}}}

\begin{document}
   \title{$\hhco$ in the Horsehead PDR: \\Photo-desorption of dust
     grain ice mantles}


   \author{V. Guzm\'{a}n\inst{1,2} \and J. Pety\inst{2,1} \and
     J.R. Goicoechea\inst{3} \and  M. Gerin\inst{1} \and
     E. Roueff\inst{4} 
         }

          \institute{LERMA - LRA, UMR 8112, Observatoire de Paris and Ecole 
            normale Sup\'{e}rieure, 24 rue Lhomond, 75231 Paris, France. \\
            \email{[viviana.guzman;maryvonne.gerin]@lra.ens.fr}
            \and
            IRAM, 300 rue de la Piscine, 38406 Grenoble Cedex, France. \\
            \email{pety@iram.fr}
            \and
            Departamento de Astrof\'{i}sica. Centro de Astrobiolog\'{i}a. 
            CSIC-INTA. Carretera de Ajalvir, Km 4. Torrej\'{o}n de Ardoz, \\
            28850 Madrid, Spain. \\
            \email{jr.goicoechea@cab.inta-csic.es}
            \and
            LUTH UMR 8102, CNRS and Observatoire de Paris, Place J. Janssen, 
            92195 Meudon Cedex, France.\\
            \email{evelyne.roueff@obspm.fr}
          }

     \date{Received May 16, 2011; accepted August 22, 2011}

 
  \abstract
   {}
   {For the first time we investigate the role of the grain
     surface chemistry in the Horsehead Photo-dissociation region (PDR).} 
   {We performed deep observations of several $\hhco$ rotational lines
     toward the PDR and its associated dense-core in the Horsehead
     nebula, where the dust is cold ($\Td \simeq 20-30$ K). We
     complemented these observations with a map of the
     $\phhco~3_{03}-2_{02}$ line at 218.2 GHz (with 12" angular
     resolution). We determine the $\hhco$ abundances using a detailed
     radiative transfer analysis and compare these results with PDR
     models that include either pure gas-phase chemistry or both
     gas-phase and grain surface chemistry.}   
   {The $\hhco$ abundances ($\simeq 2-3 \times 10^{-10}$) with
     respect to H-nuclei are similar in the PDR and dense-core. In
     the dense-core the pure gas-phase chemistry model reproduces
     the observed $\hhco$ abundance. Thus, surface processes do
     not contribute significantly to the gas-phase $\hhco$ abundance
     in the core. In contrast, the formation of $\hhco$ on
     the surface of dust grains and subsequent photo-desorption into
     the gas-phase are needed in the PDR to explain the observed gas-phase
     $\hhco$ abundance, because the gas-phase chemistry alone does not
     produce enough $\hhco$. The assignments of different
     formation routes are strengthen by the different measured 
     ortho-to-para ratio of $\hhco$: the dense-core displays the 
     equilibrium value ($\sim3$) while the PDR displays an 
     out-of-equilibrium value ($\sim2$).}
   {Photo-desorption of $\hhco$ ices is an efficient mechanism to
     release a significant amount of gas-phase $\hhco$ into the
     Horsehead PDR.} 
   

   \maketitle
%

\newcommand{\TabObsMaps}{%
  \begin{table*}
    \caption{Observation parameters for the maps shown in Fig.~\ref{fig:maps}.
      The projection center of all maps is
      $\alpha_{2000} = 05^h40^m54.27^s$, $\delta_{2000} = -02\deg 28'
      00''$.}
    \label{tab:obs:maps}
    \begin{center}
      {\tiny
        \begin{tabular}{ccrcccccccr}
          \hline \hline
          Molecule & Transition & Frequency  & Instrument & Beam   & PA     & Vel. Resol. & Int. Time &    \Tsys{} &      Noise & \multicolumn{1}{c}{Obs.date}\\
                   &            & GHz        &            & arcsec & $\deg$ & $\kms$     & hours     & K (\Tas{}) & K (\Tmb{}) & \\
          \hline
          \multicolumn{3}{c}{Continuum at 1.2\mm}                  & 30m/MAMBO & $11.7 \times 11.7$  & 0 &  --  &           -- &  -- & --         &    --  \\
          \DCOp{}  & 3-2                              & 216.112582 &  30m/HERA & $11.4 \times 11.4$  & 0 & 0.11 & 1.5/2.0$^{a}$ & 230 & 0.10       & 2006 Mar \\
          $\phhco$ & $3_{03}-2_{02}$                   & 218.222190 & 30m/HERA  & $11.9 \times 11.9$  & 0 & 0.05 & 2.1/3.4$^{a}$ & 280 & 0.32       & 2008 Jan \\
          HCO      & $1_{0,1}\,3/2, 2 - 0_{0,0}\,1/2,1$ &  86.670760 & PdBI/C\&D & $6.69 \times 4.39$ & 16 & 0.20 &     6.5$^{b}$ & 150 & 0.09$^{c}$ & 2006-2007 \\
         \hline
        \end{tabular}}
    \end{center}
    $^{a}$ Two values are given for the integration time: the on-source time 
    and the telescope time.
    $^{b}$ On-source time computed as if the source were always observed with 
    six antennae.
    $^{c}$ The noise values quoted here are the noises at the mosaic phase 
    center (mosaic noise is inhomogeneous because of the primary beam correction; it 
    steeply increases at the mosaic edges).
\end{table*}

\begin{table*}
  \caption{Observation parameters of the deep integrations of the $\ohhco$ and 
    $\phhco$ lines toward the PDR and the dense-core.}
  \begin{center}
  \begin{tabular}{llcccccccccr}
    \hline 
    \hline
    Position & Molecule  & Transition & $\nu$ & Line area & Instrument & $F_{\textrm{eff}}$ & $B_{\textrm{eff}}$ & 
    Int. Time & $T_{\textrm{peak}}$ & RMS & S/N\\
    & & & [GHz] & $\Kkms$ & & & & hours & K ($\Tmb$) & K &\\
    \hline
    & $\ohhco$ & $2_{12}-1_{11}$ & 140.839 & 1.75$\pm$0.02 & 30-m/C150 & 0.93 & 0.70 & 1.9 & 1.87 & 0.061 & 31\\
    & $\phhco$ & $2_{02}-1_{01}$ & 145.603 & 1.32$\pm$0.02 & 30-m/D150 & 0.93 & 0.69 & 1.9 & 1.61 & 0.047 & 34\\
    & $\ohhco$ & $2_{11}-1_{10}$ & 150.498 & 1.41$\pm$0.02 & 30-m/C150 & 0.93 & 0.68 & 1.4 & 1.52 & 0.053 & 29\\
PDR & $\ohhco$ & $3_{13}-2_{12}$ & 211.211 & 1.24$\pm$0.03 & 30-m/B230 & 0.91 & 0.57 & 1.1 & 1.46 & 0.096 & 15\\
    & $\phhco$ & $3_{03}-2_{02}$ & 218.222 & 0.77$\pm$0.01 & 30-m/B230 & 0.91 & 0.55 & 3.9 & 1.11 & 0.052 & 21\\
    & $\phhco$ & $3_{22}-2_{21}$ & 218.476 & 0.17$\pm$0.01 & 30-m/B230 & 0.91 & 0.55 & 2.0 & 0.27 & 0.055 & 5\\
    & $\ohhco$ & $3_{12}-2_{11}$ & 225.698 & 0.84$\pm$0.02 & 30-m/A230 & 0.91 & 0.54 & 6.5 & 1.12 & 0.079 & 14\\
    \hline                                                           
    & $\ohhco$ & $2_{12}-1_{11}$ & 140.839 & 2.56$\pm$0.01 & 30-m/C150 & 0.93 & 0.70 & 3.7 & 3.46 & 0.036 & 96\\
    & $\phhco$ & $2_{02}-1_{01}$ & 145.603 & 1.75$\pm$0.02 & 30-m/D150 & 0.93 & 0.69 & 1.9 & 2.62 & 0.044 & 60\\
    & $\ohhco$ & $2_{11}-1_{10}$ & 150.498 & 1.89$\pm$0.01 & 30-m/C150 & 0.93 & 0.68 & 1.5 & 2.52 & 0.052 & 49\\
    & $\ohhco$ & $3_{13}-2_{12}$ & 211.211 & 1.93$\pm$0.02 & 30-m/B230 & 0.91 & 0.57 & 2.0 & 3.02 & 0.065 & 47\\
    & $\phhco$ & $3_{03}-2_{02}$ & 218.222 & 1.03$\pm$0.01 & 30-m/B230 & 0.91 & 0.55 & 3.0 & 1.83 & 0.057 & 32\\
    & $\phhco$ & $3_{22}-2_{21}$ & 218.476 & 0.04$\pm$0.01 & 30-m/B230 & 0.91 & 0.55 & 4.5 & 0.06 & 0.037 & 2\\
    Dense-core & $\ohhco$ &  $3_{12}-2_{11}$ & 225.698 & 1.27$\pm$0.02& 30-m/A230 & 0.91 & 0.54 & 8.4 & 1.96 & 0.073 & 27\\  
    & $\ohhthico$ & $2_{12}-1_{11}$ & 137.450 & 0.09$\pm$0.02 & 30-m/D150 & 0.95 & 0.70 & 2.0 & 0.11 & 0.063& 2\\
    & $\phhthico$ & $2_{02}-1_{01}$ & 141.984 & 0.10$\pm$0.01 & 30-m/D150 & 0.95 & 0.70 & 1.5 & 0.11 & 0.060& 2\\
    & HDCO & $2_{11}-1_{10}$ & 134.285 & 0.13$\pm$0.01 & 30-m/C150 & 0.94 & 0.71 & 2.0 & 0.32 & 0.042 & 8\\     
    & HDCO & $3_{12}-2_{11}$ & 201.341 & 0.05$\pm$0.01 & 30-m/A230 & 0.91 & 0.59 & 3.5 & 0.13 & 0.032 & 4\\   
    & $\pddco$ & $2_{12}-1_{11}$ & 110.838 & 0.04$\pm$0.01 & 30-m/A100 & 0.95 & 0.75 & 4.9 & 0.08 & 0.031 & 3\\
    & $\oddco$ & $4_{04}-3_{03}$ & 231.410 & 0.04$\pm$0.01 & 30-m/A230 & 0.91 & 0.53 & 4.5 & 0.09 & 0.068 & 1\\
    \hline
  \end{tabular}
  \end{center}
  \label{tab:obs:lines}
\end{table*}
}

\newcommand{\FigMaps}{%
\begin{figure}[t!]
  \centering %
  \includegraphics[width=\hsize{}]{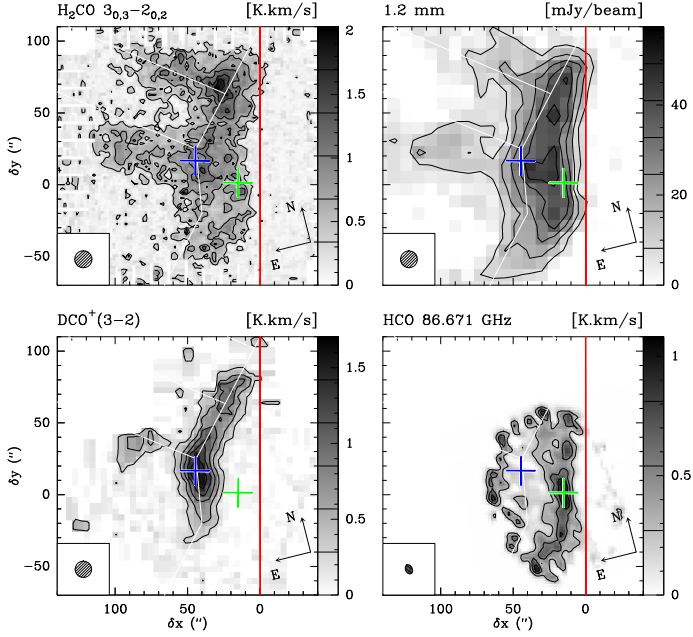} %
   \caption{Integrated intensity maps of the Horsehead edge. The intensities 
     are expressed in the main-beam temperature scale. Maps were
     rotated by 14\deg{} counter--clockwise around the projection
     center, located at $(\delta x,\delta y)$ = $(20'',0'')$, to bring
     the exciting star direction in the horizontal direction and the
     horizontal zero was set at the PDR edge, delineated by the
     red vertical line. The crosses show the positions of the PDR
     (green) and the dense-core (blue), where deep integrations
     were performed at IRAM-30m (see
     Fig.~\ref{fig:montecarlo}). The spatial resolution is plotted in
     the bottom left corner. Values of contour levels are shown on
     each image lookup table. The emission of all lines is integrated
     between 10.1 and $11.1\kms$.} 
  \label{fig:maps}
\end{figure}
}

\newcommand{\SpecParam}{%
\begin{table}[t!]
  \caption{Spectroscopic parameters of the observed lines obtained from the CDMS data base \citep{muller01}. }
  \begin{center}
  \begin{tabular}{rccrcrccc}
    \hline \hline
    Molecule & Transition & $\nu$ & $E_u$ & $A_{ul}$ & $g_u$ \\
    & & [GHz] & [K] & [s$^{-1}$] & \\
   \hline
    $\ohhco$    & $2_{12}-1_{11}$ & 140.839 & 21.92 & 5.3$\times 10^{-5}$ & 15 \\
    $\phhco$    & $2_{02}-1_{01}$ & 145.603 & 10.48 & 7.8$\times 10^{-5}$ & 5  \\
    $\ohhco$    & $2_{11}-1_{10}$ & 150.498 & 22.62 & 6.5$\times 10^{-5}$ & 15 \\
    $\ohhco$    & $3_{13}-2_{12}$ & 211.211 & 32.06 & 2.3$\times 10^{-4}$ & 21 \\
    $\phhco$    & $3_{03}-2_{02}$ & 218.222 & 20.96 & 2.8$\times 10^{-4}$ & 7  \\
    $\phhco$    & $3_{22}-2_{21}$ & 218.476 & 68.09 & 1.6$\times 10^{-4}$ & 7  \\
    $\ohhco$    & $3_{12}-2_{11}$ & 225.698 & 33.45 & 2.8$\times 10^{-4}$ & 21 \\
    \hline                                  
    $\ohhthico$ & $2_{12}-1_{11}$ & 137.450 & 10.51 & 4.9$\times 10^{-5}$ & 15 \\
    $\phhthico$ & $2_{02}-1_{01}$ & 141.984 & 2.37  & 7.2$\times 10^{-5}$ & 5  \\
    \hline                                  
    HDCO         & $2_{11}-1_{10}$ & 134.285 & 17.63 & 4.6$\times 10^{-5}$ & 5 \\
    HDCO         & $3_{12}-2_{11}$ & 201.341 & 27.29 & 2.0$\times 10^{-4}$ & 7 \\
    \hline                                  
    $\oddco$    & $2_{12}-1_{11}$ & 110.838 & 13.37 & 2.6$\times 10^{-5}$ & 5  \\
    $\pddco$    & $4_{04}-3_{03}$ & 231.410 & 27.88 & 3.5$\times 10^{-4}$ & 18 \\
    \hline
  \end{tabular}
  \end{center}
  \label{tab:obs_par}
\end{table}

\begin{figure}[h!]
  \begin{center}
    \includegraphics[scale=0.115,angle=-90]{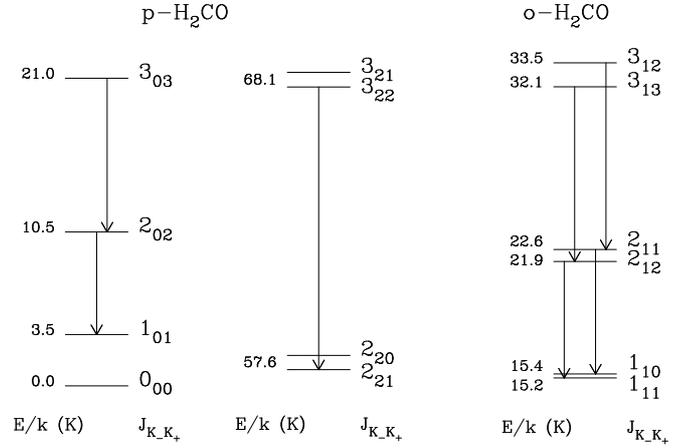}
  \end{center}
  \caption{Lower energy rotational levels of para- (left) and ortho-$\hhco$
    (right). The energy above para ground-state is shown at the left of each
    level. The arrows indicate the transitions detected in the
    Horsehead.}
  \label{fig:level_diag}
\end{figure}
}

\newcommand{\FigMontecarlo}{%
  \begin{figure*}[t!]
    \centering
    \includegraphics[scale=.6]{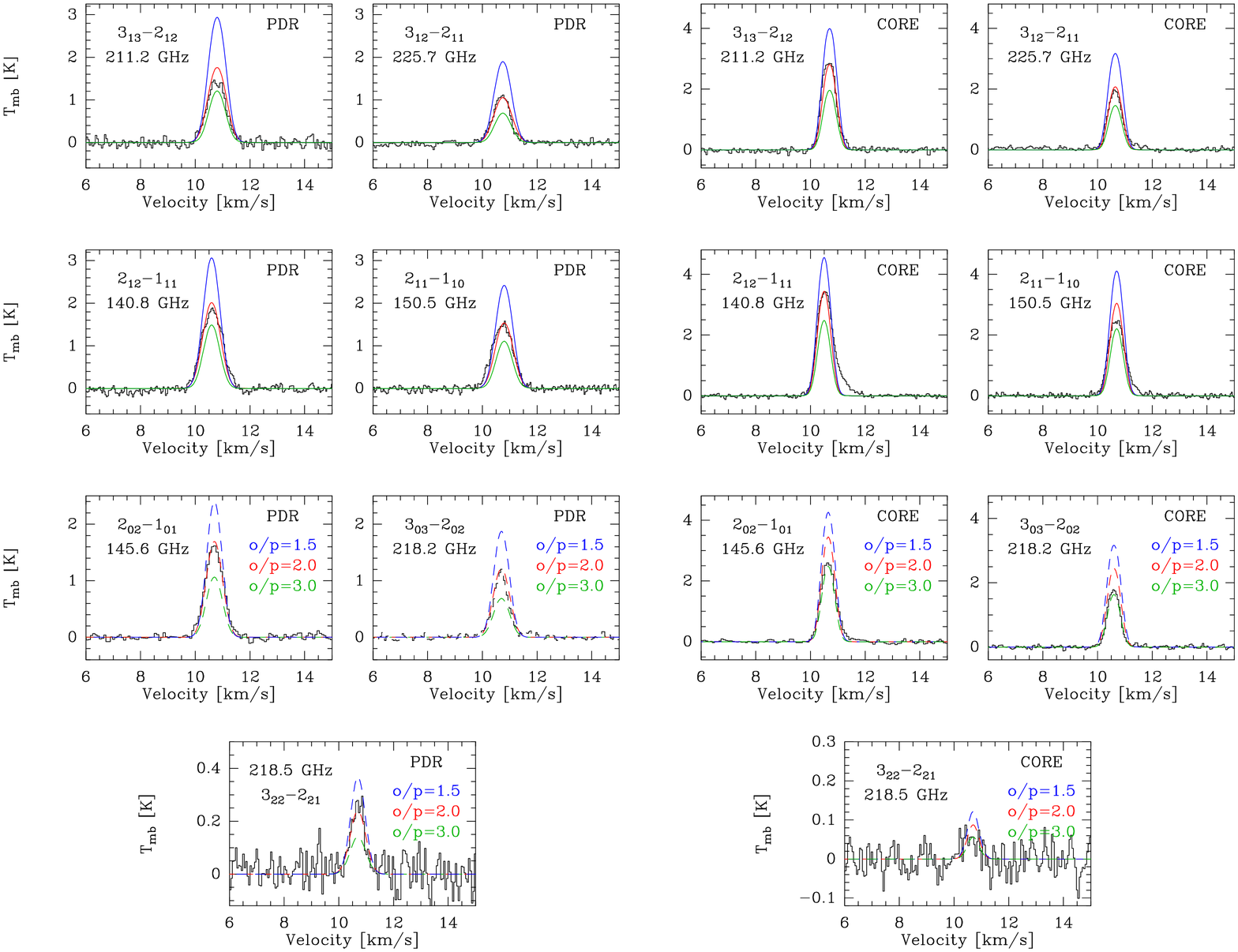}
    \caption[Optional caption for list of figures]{Radiative transfer modeling 
      of $\hhco$ lines for two positions toward the Horsehead. Two left 
      columns: the PDR position ($\Tkin=60$ K, $n(\hh)~=~6~\times~10^4 \pccm$, 
      N($\ohhco$)~=~$7.2~\times~10^{12} \pscm$) and two right columns: the 
      dense-core position ( $\Tkin=20$ K, $n(\hh)~=~10^5 \pccm$, 
      N($\ohhco$)~=~$9.6~\times~10^{12} \pscm$). The two top rows display the 
      ortho lines, for which we varied the column density around the best 
      match (red curve) by a factor of 1.5 (blue curve) and 1/1.5 (green 
      curve). The two bottom rows display the para lines, for which we kept 
      the column density of the best match for $\ohhco$ (red curves) constant  
      and varied the ortho-to-para ratio of $\hhco$: o/p = 1.5 (dashed blue), 
      o/p=2 (dashed red) and o/p=3 (dashed green)}
    \label{fig:montecarlo}
  \end{figure*}
}

\newcommand{\FigDeuterated}{%
\begin{figure}[t!]
  \centering
    \includegraphics[scale=0.55,angle=-90]{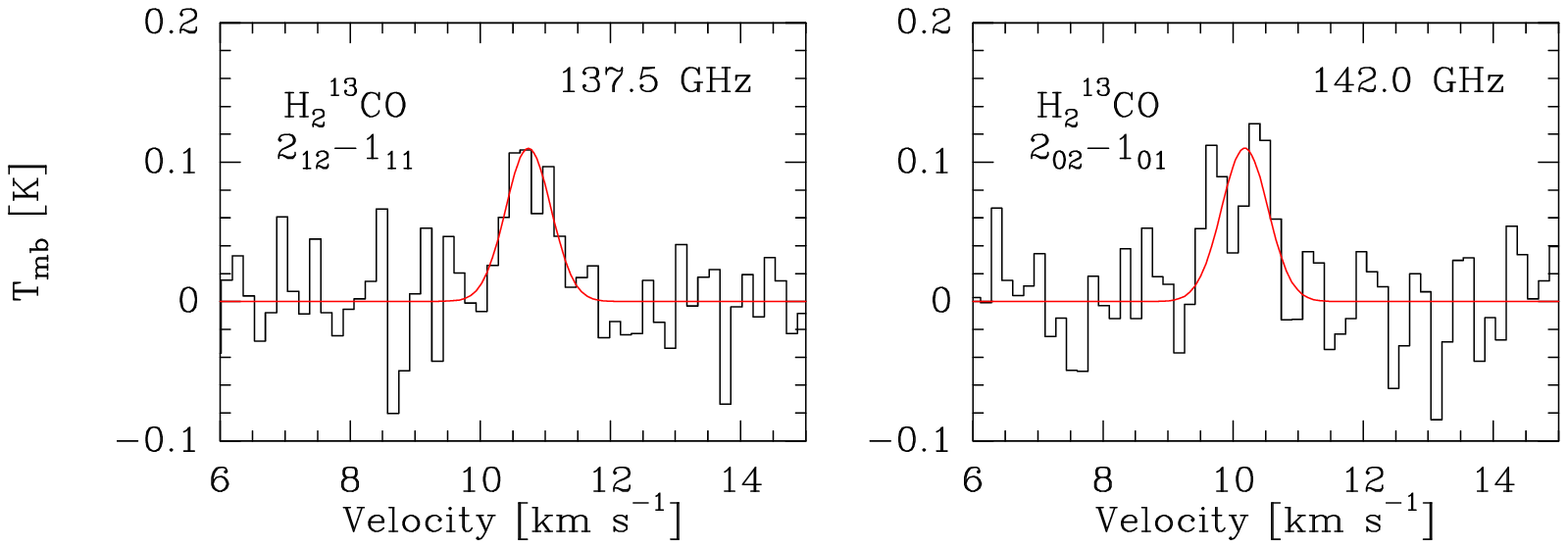} \\
    \vspace{0.2cm}
    \includegraphics[scale=.55,angle=-90]{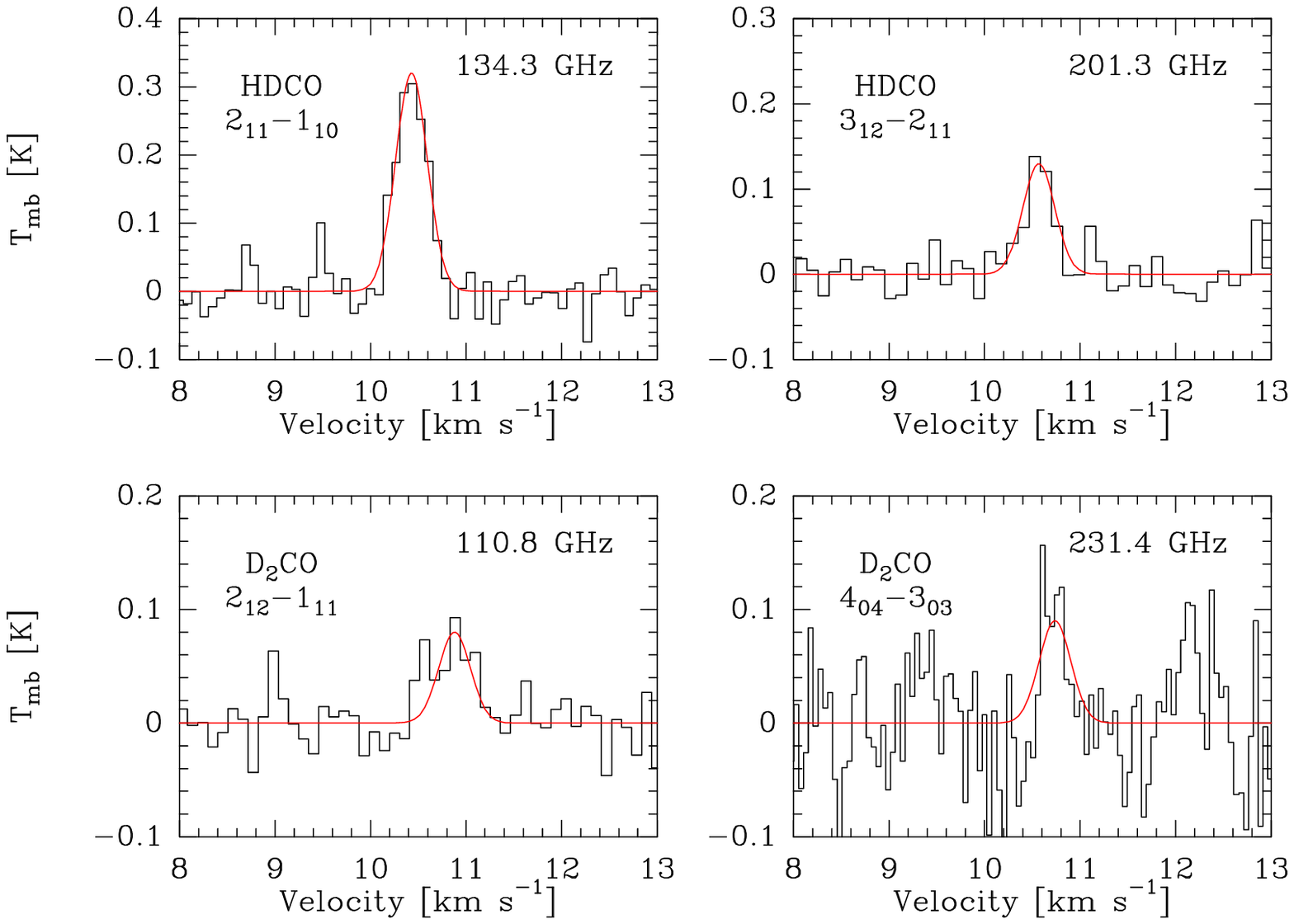}
  \caption{$\hhthico$ and deuterated $\hhco$ lines detected toward the 
    dense-core. Gaussian fits are shown with red lines. For HDCO and 
    $\ddco$ the line width was fixed to the width of the HDCO ($2_{11}-1_{10}$) 
    line, because it has the best signal-to-noise ratio. }
  \label{fig:h213co_deuterated}
\end{figure}
}

\newcommand{\FigRotDiag}{%
\begin{figure}[t!]
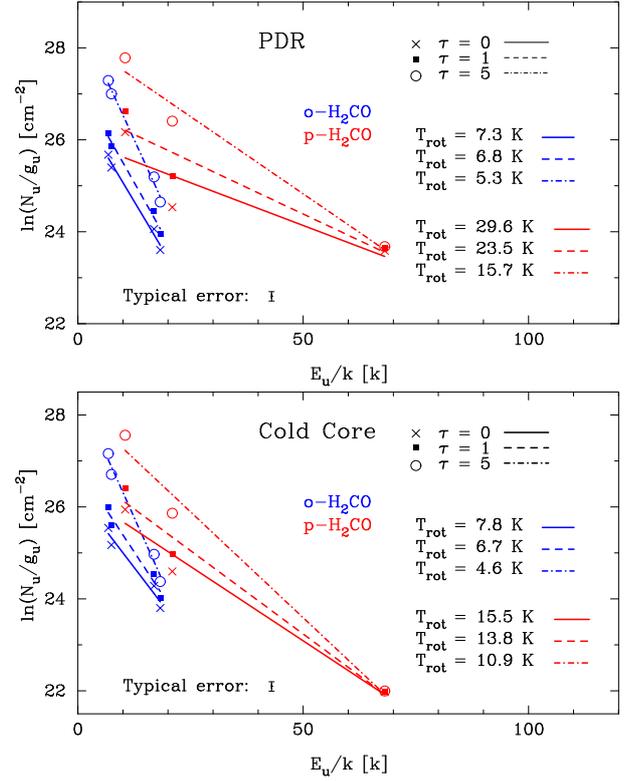

  \centering
 \includegraphics[scale=0.8,angle=-90]{rotation_diagram_H2CO_PDR.ps}
  \vspace{0.1cm}
  \includegraphics[scale=0.8,angle=-90]{rotation_diagram_H2CO_CC.ps}
  \caption{$\hhco$ rotational diagrams corrected for line-opacity effects at 
    the PDR and dense-core position. Rotational temperatures are shown for each 
    considered opacity.}
  \label{fig:rot_diags}
\end{figure}
}

\newcommand{\TabCritDens}{%
\begin{table}[t!]
  \caption{$\hhco$ critical densities ($\pccm$) for three different
    colliding partners computed for $\Tkin = 60$ K.} 
  \centering
  \begin{tabular}{c|ccc}
    \hline \hline
    $J_{K_a K_c}$ & p-H$_2$          & o-H$_2$           & He \\
    \hline
    $2_{02}$ & $7.2\times10^5$ & $3.6\times10^5$ & $1.3\times10^6$ \\
    $3_{03}$ & $1.6\times10^6$ & $9.9\times10^5$ & $4.2\times10^6$ \\
    $3_{22}$ & $5.8\times10^5$ & $4.7\times10^5$ & $2.5\times10^6$ \\
    $2_{12}$ & $3.7\times10^5$ & $2.5\times10^5$ & $8.1\times10^5$ \\
    $2_{11}$ & $4.3\times10^5$ & $2.2\times10^5$ & $8.7\times10^5$ \\
    $3_{13}$ & $9.7\times10^5$ & $7.0\times10^5$ & $2.3\times10^6$ \\
    $3_{12}$ & $1.3\times10^6$ & $7.9\times10^5$ & $3.2\times10^6$ \\
    \hline 
  \end{tabular}
  \label{tab:ncrit}
\end{table}
}

\newcommand{\TabAbund}{%
\begin{table}[t!]
  \centering
  \caption{Column densities and abundances.}
  \begin{threeparttable}[b]
  \begin{tabular}{clcc}
    \hline 
    \hline
   & Molecule             &   PDR             & Dense-core         \\
   \hline
    &$N_{\mathrm{H}}$ & $3.8\times10^{22}$ & $6.4\times10^{22}$ \\
    &$N(\ohhco$)    & $7.2\times10^{12}$ & $9.6\times10^{12}$  \\
 \multirow{2}{*}{Column densities}  &$N(\phhco$)    & $3.6\times10^{12}$ & $3.2\times10^{12}$  \\
    \multirow{2}{*}{$[\pscm]$} & $N$(HCO)      \tnote{a} & $3.2\times10^{13}$ & $<4.6\times10^{12}$ \\ 
    &$N$(HDCO)     \tnote{b} &        -          & $1.6\times10^{12}$ \\
    &$N(\oddco$)   \tnote{b} &        -          & $5.1\times10^{11}$ \\
    &$N(\pddco$)   \tnote{b} &        -          & $5.1\times10^{11}$ \\
    \hline
    &$[\ohhco]$     & $1.9\times10^{-10}$ & $1.5\times10^{-10}$ \\
    \multirow{2}{*}{Abundances} &$[\phhco]$     & $9.5\times10^{-11}$ & $5.0\times10^{-11}$ \\
    \multirow{2}{*}{$[\x] = \frac{N(\x)}{(N(\h)+2~N(\hh))}$} & $[$HCO$]$      & $8.4\times10^{-10}$  & $<7.2\times10^{-11}$ \\
    &$[$HDCO$]$     &         -          & $2.5\times10^{-11}$ \\
    &$[\ddco]$      &         -          & $1.6\times10^{-11}$ \\
    \hline
  \end{tabular}
  \begin{tablenotes}
    \item[a]{\cite{gerin09}}
    \item[b]{For $\Tex = 6$ K (LTE).}
  \end{tablenotes}  
  \end{threeparttable}
  \label{tab:column_densities}
\end{table}
}

\newcommand{\FigMeudon}{%
\begin{figure}[t!]
  \centering
\includegraphics[scale=0.8,angle=0]{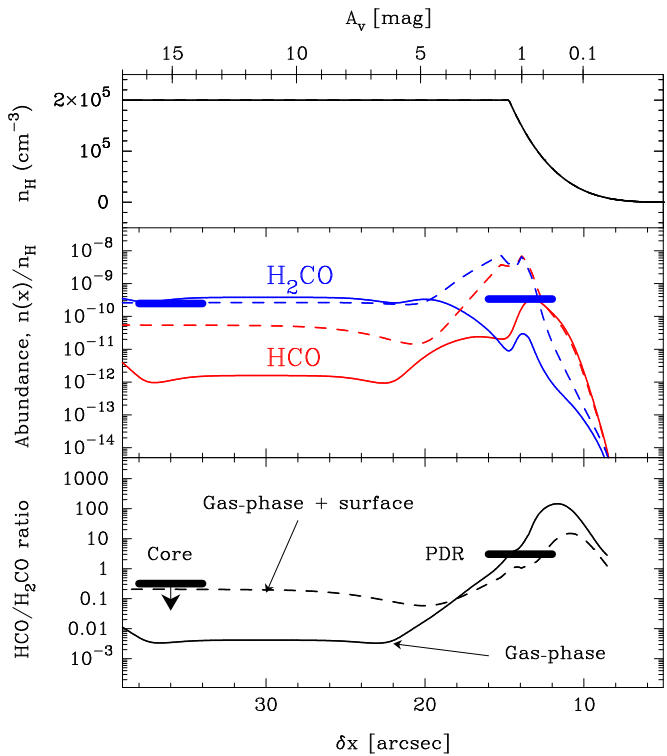} 
  \caption{Photochemical model of the Horsehead PDR. \textit{Upper
      panel:} PDR density profile ($\nH = n($H$) + 2 n($H$_2)$ in
    $\pccm$). \textit{Middle panel:} Predicted abundance (relative to  
    $\nH$) of $\hhco$ (blue) and HCO (red). \textit{Lower
      panel:} Predicted HCO/$\hhco$ abundance ratio. In the two
      bottom panels, models shown as solid lines include pure
      gas-phase chemistry and models shown as dashed lines include
      gas-phase as well as grain surface chemistry. The horizontal bars 
      show the measured $\hhco$ abundances and abundance ratios. }    
    \label{fig:meudon}
\end{figure}
}


\TabObsMaps{} %

\section{Introduction}

Photo-dissociation region (PDR) models are used to understand the
evolution of the far-UV illuminated matter both in our Galaxy and in
external galaxies. The spectacular instrumental improvements, which
happen in radioastronomy with the advent of Herschel, ALMA and NOEMA,
call for matching progresses in PDR modeling. In particular, the
physics and chemistry of the dust grains and of the gas-phase are
intricately intertwined. It is well known that the formation of ice
grain mantles leads to the removal of chemical reservoirs like CO, O,
and other abundant species from the gas phase, enabling new chemical
routes to be opened and others to be closed. Despite their low
temperature, the mantles are chemically active. 
Hydrogenation/deuteration reactions are known to be efficient,   
because hydrogen (or deuterium atoms) can migrate on the surfaces, but
reactions with O, N, and C must also be considered. Complex molecules
may therefore be formed before they are released into the gas phase. Moreover,
the release of the products into the gas phase happens either through
thermal processes (evaporation) or non-thermal ones
(cosmic ray or far-UV photon-induced desorption). Recent
photo-desorption experiments on water and CO ices show that this mechanism is
much more efficient than previously thought
\citep{oberg09b,oberg09a,munoz-caro10}. These results
led various groups to include photo-desorption into PDR models
\citep[see the results on $\hho$ and O$_2$
by][]{hollenbach09,walsh10,hassel10}. The availability of 
well-defined observations is essential here to distinguish between
chemical assumptions about the significant grain surface processes,
\ie{}, adsorption, desorption, and diffusion. It is now confirmed that
some interstellar species are mostly formed in the gas-phase (CO for
instance), others on grains \citep[CH$_3$OH,][]{garrod07}, while the
chemical routes for other complex species such as formaldehyde, are
still debated because it is likely that solid and gas-phase processes
are both needed.   

Formaldehyde ($\hhco$) was the first organic molecule
discovered in the interstellar medium \citep{snyder69}. It is a
relatively simple organic molecule that can be formed in the
gas-phase and on the surface of dust grains. In the warm gas, $\hhco$ 
can trigger the formation of more complex organic molecules
\citep{charnley92}. It is one of the most popular  
molecules used for studying the physical conditions of the gas in
astrophysical sources. Indeed, it is a good probe of the temperature
and density of the gas~\citep{mangum93}. Owing to its large dipole
moment (2.3~Debye), its rotational lines are easy to detect from
ground-based observations. It is present in a variety of environments,
such as Galactic HII regions \citep[\eg,][]{downes80}, proto-stellar
cores \citep[\eg,][]{young04,maret04}, young stellar objects
\citep[\eg,][]{araya07}, PDRs \citep[\eg,][]{leurini10}, starburst
galaxies \citep[\eg,][]{mangum08} and comets
\citep[\eg,][]{snyder89,milam06}. 

The Horsehead PDR is particularly well-suited to investigate grain surface 
chemistry in a UV irradiated environment. It is viewed nearly edge-on
\citep{habart05} at a distance of 400~pc (implying that 10''
correspond to 0.02~pc). Thus, it is possible to study the interaction
of far-UV radiation with dense interstellar clouds and the transition
from warm to cold gas in a PDR with a simple geometry, very close to
the prototypical kind of source needed to serve as a reference to
chemical models. Its relatively low UV illumination \citep[$\chi = 60$
in Draine units,][]{draine78} and high density ($\nH\sim10^4-10^5 \pccm$)
implies low dust grain temperatures, from $\Td \sim30$~K in the PDR to
$\Td \sim20$~K deeper inside the cloud \citep{goicoechea09a}. The
release of the grain mantle products into the gas phase is consequently 
dominated by photo-desorption, while regions with warmer dust (the Orion 
bar or the star-forming cores) provide mixed information on the 
thermal and non-thermal processes \citep[\eg,][]{Bisschop07}.  

In this paper we present deep observations of several formaldehyde
lines toward two particular positions in the Horsehead nebula: the
PDR, corresponding to the peak of the HCO emission \citep{gerin09},
where the gas is warm ($\Tkin \sim 60$ K); and the dense-core, 
a cold ($\Tkin \le 20$ K) condensation located less than 40''
away from the PDR edge, where HCO$^+$ is highly deuterated
\citep{pety07}. We present the observations and data reduction in
Sect. \ref{sec:obs}, while the results and abundances are given in
Sect. \ref{sec:results}. In Sect. \ref{sec:chemistry} we present the 
$\hhco$ chemistry and PDR modeling. A discussion is given in Sect.  
\ref{sec:discussion} and we conclude in Sect. \ref{sec:conclusions}.  

\FigMaps{} %

\section{Observations and data reduction}
\label{sec:obs}

Tables~\ref{tab:obs:maps} and~\ref{tab:obs:lines} summarize the
observation parameters for the data obtained with the IRAM-30m and
PdBI telescopes. Fig.~\ref{fig:maps} displays the $\phhco$,
  HCO, \DCOp{} and 1.2\mm{} continuum maps. The $\phhco$ line was
mapped during 3.3 hours of good winter weather ($\sim 1\mm$ of water
vapor) using the first polarization (\ie{} nine of the eighteen
available pixels) of the IRAM-30m/HERA single-sideband multi-beam
receiver. We used the frequency-switched, on-the-fly observing
mode. We observed along and perpendicular to the direction of the
exciting star in zigzags (\ie{} $\pm$ the lambda and beta scanning
direction). The multi-beam system was rotated by $9.6\deg$ with
respect to the scanning direction. This ensured Nyquist sampling
between the rows except at the edges of the map. The fully sampled
part of the map covered a $150'' \times 150''$ portion of the sky. A
detailed description of the HCO, $\DCOp$ and 1.2 mm continuum
observations and data reductions can be found in 
~\citet{gerin09}, ~\citet{pety07}, and~\citet{hilyblant05}. 

We performed deep integrations of $\ohhco$ and $\phhco$ 
low-energy rotational lines (see Fig.~\ref{fig:level_diag} and
\ref{fig:montecarlo}) centered on the PDR
and the dense-core. To obtain these deep integration spectra, we used
the position-switching observing mode. The on-off cycle duration was 1
minute and the off-position offsets were ($\delta$ RA, $\delta$ Dec) = 
(-100'',0''), \ie{} the {\sc{H\,ii}} region ionized by $\sigma$Ori
and free of molecular emission. From our knowledge of the IRAM-30m
telescope we estimate the absolute position accuracy to be $3''$. 

\SpecParam{} %

The data processing was made with the \GILDAS{}\footnote{See
  \texttt{http://www.iram.fr/IRAMFR/GILDAS} for more information about the
  \GILDAS{} softwares.} softwares~\citep{pety05}.  The IRAM-30m data were 
first calibrated to the \Tas{} scale using the chopper-wheel 
method~\citep{penzias73}, and finally converted to main-beam temperatures 
(\Tmb{}) using the forward and main-beam efficiencies (\Feff{} \& \Beff{})
displayed in Table~\ref{tab:obs:lines}. The resulting 
amplitude accuracy is $\sim 10\%$.  Frequency-switched spectra were folded 
using the standard shift-and-add method before baseline subtraction. The 
resulting spectra were finally gridded through convolution with a Gaussian to 
obtain the maps.


\section{Results}
\label{sec:results}

\FigMontecarlo{} %

\subsection{$\hhco$ spatial distribution}

The 218.2 GHz $\phhco$ integrated line-intensity map is shown in
Fig.~\ref{fig:maps} together with the 86.7 GHz HCO, 216.1 GHz $\DCOp$
integrated line-intensity maps and the 1.2 mm continuum-emission
map. Formaldehyde emission is extended throughout the Horsehead with
a relatively constant intensity. The $\hhco$ spatial
distribution ressembles the 1.2~mm continuum emission: It follows the
top of the famous Horsehead nebula from its front to its mane. It
also delineates the throat of the Horsehead. The peak of the $\hhco$
emission spatially coincides with the peak of the $\DCOp$ emission,
which arises from a cold dense-core. However, $\hhco$
emission is also clearly present along the PDR, which is traced by
the HCO emission. The PDR and dense-core, namely the peaks of the
HCO and $\DCOp$ emission are shown with green and blue crosses
respectively. 
Gaussian fits of the $\hhco$ lines at the HCO peak result in broader  
line widths than at the $\DCOp$ peak. That the lines are broader 
in the PDR confirms that $\hhco$ lines toward the $\DCOp$ peak arise 
from the dense-core rather than from the illuminated surface of the 
cloud. There is a peak in the $\hhco$ emission toward the north-west 
region of the nebula, near the edge of the PDR, where two protostars 
have been identified \cite[B33-1 and B33-28,][]{bowler09}. These 
protostars heat the dust around them, so it is likely that $\hhco$ 
has been evaporated from the grain ice mantles.    

\subsection{$\hhco$ column density}

We computed the column densities of $\hhco$ at the PDR and the dense-core 
positions. For this we first used the $\hhthico$ lines to estimate
the optical depth of the $\hhco$ lines. Then, we made a first
estimate of the column densities and excitation temperatures using
rotational diagrams. Finally, we used these first estimates as an
input for a detailed nonlocal non-LTE excitation and radiative
transfer analysis to compute the $\hhco$ abundances. The spectroscopic
parameters for the detected transitions (shown in
Fig.~\ref{fig:level_diag}) are given in Table~\ref{tab:obs_par}. We
assumed that the emission is extended and fills the 30m beam, as shown
by the map of the $3_{03}-2_{02}$ transition (see
Fig.~\ref{fig:maps}).   

\subsubsection{Opacity of the $\hhco$ lines}

We detected two transitions of the formaldehyde isotopologue
$\hhthico$ in the dense-core position (see upper panels in
Fig.~\ref{fig:h213co_deuterated}). By comparing the flux between  
$\hhco$ and $\hhthico$ for the same transition it is possible to
estimate the opacity of the $\hhco$ line, assuming that the $\hhthico$
line is optically thin, as follows: 
\begin{equation} 
  \frac{F_{\hhco}}{F_{\hhthico}} = \frac{[^{12}\textrm{C}]}{[^{13}\textrm{C}]} 
\beta
\end{equation}
where $\beta$ is the escape probability function, which in the case of a
homogeneous slab of gas \citep{dejong80} is equal to
\begin{equation} 
  \beta = \frac{1-\exp(-3\tau)}{3\tau}
\end{equation}
The isotopic abundance ratio $^{12}$C/$^{13}$C $\simeq$ 60
\citep{langer90,savage02} is almost twice the line intensity ratio between  
formaldehyde and its isotopologue, and therefore the $\hhco$ lines
have moderate opacities. From the observations we estimate
$\tau_{2_{12}-2_{11}} \sim 1.6$ and $\tau_{2_{02}-1_{01}} \sim 1.9$ for
H$_2$CO in the dense-core. 

\FigDeuterated{}%

\subsubsection{Rotational diagram analysis}

First-order estimates of the beam-averaged column densities and the rotational 
temperatures can be found by means of the widely used rotational diagram 
analysis~\citep{goldsmith99}. To do this, we assume that the gas is
under LTE, and therefore all excitation temperatures are the same, and
the energy levels are populated following Boltzmann's law. We built
rotational diagrams corrected for line-opacity effects through  
\begin{equation}
  \ln \frac{N^{\mathrm{thin}}_u}{g_u} + \ln C_{\tau} = \ln \frac{N}{Z} - 
  \frac{E_u}{k\Trot},
\end{equation}
where $N$ is the total column density of the molecule, $g_u$ is the level 
degeneracy, $E_u/k$ is the energy of the upper level in K, $Z$ is the partition 
function at the rotational temperature \Trot{}, 
$C_{\tau} = \frac{\tau}{1-e^{-\tau}} \leq 1$ is a line-opacity correction
factor, where $\tau$ is the opacity of the line, and
$N^{\mathrm{thin}}_u$ is the column density of the upper level for an
optically thin line when the source fills the beam. This last
parameter is given by
\begin{equation}
  N^{\mathrm{thin}}_u = \frac{8 \pi k \nu^2 W}{h c^3 A_{ul}}, 
\end{equation}
where $k$ is the Boltzmann constant, $\nu$ is the line frequency, $W$ is the 
integrated line intensity, $h$ is the Planck constant, $c$ is the speed of 
light and $A_{ul}$ is the Einstein coefficient for spontaneous emission.

Ortho- and para forms of $\hhco$ are treated as different species because 
radiative transitions between them are forbidden. Resulting rotational 
diagrams are shown in Fig.~\ref{fig:rot_diags} for three different
$\ohhco~(2_{12}-1_{11})$ and $\phhco~(2_{02}-1_{01})$ line-opacities ($\tau
= 0,1$ and 5). We find column densities of $N \sim 10^{12} -10^{13}
\pscm$, depending on the opacity. We infer very different
rotational temperatures for $\ohhco$ ($\Trot \sim 4-8$ K) and $\phhco$
($\Trot \sim 10-30$ K), which are also lower than the well-known
conditions in the PDR ($\Tkin \sim 60$ K) and in the dense-core
($\Tkin \sim 20$ K). This suggests that the gas is far from
thermalization, and therefore we used these column densities and
rotational temperatures as an input for a more complex analysis to
derive the H$_2$CO column densities. 

\FigRotDiag{}%

\subsubsection{Radiative transfer models}

The critical density of a given collisional partner corresponds to the density
at which the sum of spontaneous radiative de-excitation rates is equal to the 
sum of collisional de-excitation rates ($\gamma$) of a given level

\TabCritDens{}%

\begin{equation}
  n_{cr}(J_{K_a K_c}, \Tkin) = \frac{\sum_{J^{\prime}_{K^{\prime}_a
  K^{\prime}_c}} A(J_{K_a K_c} \rightarrow J^{\prime}_{K^{\prime}_a
  K^{\prime}_c})}{\sum_{J^{\prime}_{K^{\prime}_a K^{\prime}_c}} \gamma(J_{K_a K_c}
  \rightarrow J^{\prime}_{K^{\prime}_a K^{\prime}_c}, \Tkin)}, 
\end{equation}

Formaldehyde lines have high critical densities ($\sim10^6\pccm$, see
Table~\ref{tab:ncrit}) compared to the H$_2$ density in the Horsehead 
($\sim~10^4-10^5\pccm$). Because we expect subthermal 
emission ($\Tex~\ll~\Tkin$) for transitions with high critical densities 
compared to the H$_2$ density, we used a nonlocal non-LTE
radiative transfer code adapted to the Horsehead geometry to model
the observed $\hhco$ line intensities \citep{goicoechea06}. We used
a nonlocal code to take into account the radiative
coupling between different cloud positions that might affect the 
population of the energy levels. The code is able to predict the
line profiles. It takes into account line trapping, collisional
excitation and  radiative excitation by absorption of cosmic microwave
background and dust continuum photons. We included 40 rotational
levels for $\ohhco$ and 41 rotational levels for $\phhco$, where the
maximum energy level lies at $\sim 285$ K for both species. We
considered $\ohh$, $\phh$ and He as collision partners with the
following collisional excitation rates: 
\begin{itemize}
\item Collisional rates of $\ohhco$ and $\phhco$ with He are taken from 
\citet{green91}. 
\item Collisional rates of $\ohhco$ with $\ohh$ and $\phh$ from 
  \citet{troscompt09} for the first 10 energy levels, \ie{}
  $E_u\leq50$ K. We complemented these data with He collision rates of 
  \citet{green91} scaled to $\hh$. Following the new $\hhco-$$\hh$
  collisional rate calculations, we scaled the $\hhco-$He rates by a
  factor 2.5 instead of the usual $\sim1.4$ mass factor (A. Faure
  priv. communication).  
\item Collisional rates of $\phhco$ with $\ohh$ and $\phh$ from
  Troscompt et al., to be submitted.
\end{itemize}
 
Results are presented in Fig.~\ref{fig:montecarlo} for three different 
column densities. Best matches (see Table~\ref{tab:column_densities})
are for column densities of N($\ohhco$) = $7.2~\times~10^{12} \pscm$ and
N($\phhco$) = $3.6~\times~10^{12} \pscm$ in the PDR position, and
N($\ohhco$) = $9.6~\times~10^{12} \pscm$ and N($\phhco$) =
$3.2~\times~10^{12} \pscm$ in the dense-core position. In the
excitation- and radiative transfer models we adopt 
an $\hh$ ortho-to-para ratio of 3 (high-temperature limit),
although it is likely that the ortho-to-para ratio is lower in the
Horsehead \citep[\eg,][]{habart11}. Indeed, the $\hhco$ column
densities are not sensitive to the change of the $\hh$ ortho-to-para
ratio for the physical conditions of the Horsehead (see Appendix
\ref{app:1}).  

\subsection{$\hhco$ ortho-to-para ratio}

The ratio between the column densities of $\ohhco$ and $\phhco$ provides 
information about the formation of the molecule, because the characteristic 
conversion time from one symmetry state to the other is longer than the
$\hhco$ lifetime \citep{tudorie06}. When the molecule forms in the gas-phase, a
ratio of 3 is expected, which corresponds to the statistical weight ratio
between the ground states of the ortho- and para species. A ratio lower than 3
is expected when the molecule is formed on the surface of cold ($\Td
\lesssim 20$~K) dust grains \citep{kahane84,dickens99}. From the
derived column densities we infer $\hhco$ ortho-to-para ratios of
$\sim2$ in the PDR and of $\sim3$ in the dense-core. This suggests that 
in the dense-core $\hhco$ is mainly formed in the gas-phase, whereas 
in the PDR $\hhco$ is formed on the surface of dust grains.
\cite{dickens99} measured $\hhco$ ortho-to-para ratios between 1.5
and 2 toward star-forming cores with outflows, and ratios near 3
toward three quiescent cores. \cite{jorgensen05} also found an
ortho-to-para ratio of 1.6 in the envelopes around low-mass protostars.   

\TabAbund{}%

\subsection{HDCO and $\ddco$ column densities}

We detected HDCO and $\ddco$ in the dense-core (see two bottom rows in 
Fig.~\ref{fig:h213co_deuterated}), and we estimated their abundances assuming 
LTE. For $\Tex = 6$~K we obtain N(HDCO) = $1.6\times10^{12} \pscm$, 
N($\ddco$) = $5.1\times10^{11} \pscm$ and a $\ddco$ ortho-to-para ratio of 1, 
which translates into relative abundances or fractionation levels 
[HDCO]/$[\hhco]$ = 0.11 and $[\ddco]/[\hhco]$ = 0.04 for the inferred 
formaldehyde column densities in the dense-core.

Deuterium fractionation can occur in the gas-phase by means of ion-molecule 
reactions, where D is transferred from HD to other species. High abundances of 
deuterated molecules compared to the elemental D/H abundance \citep[$\sim1.5 
\times 10^{-5}$,][]{linsky06} have been observed in different astrophysical 
environments, from cold dense cores and hot molecular cores even to PDRs. 
\cite{pety07} found high deuteration ($[\DCOp]/$[HCO$^+]$ $>0.02$) in the 
Horsehead dense-core. A pure gas-phase chemical model was able to reproduce 
the observed fractionation level of HCO$^+$ for $\Tkin \le 20$ K. 
\cite{parise09} found high fractionation levels for DCN and HDCO toward the 
Orion Bar PDR ([XD]/[XH] $\sim 0.01$). They found that these ratios are 
consistent with pure gas-phase chemistry models where the gas is warm 
($>50$ K), so the deuterium chemistry is driven mainly by CH$_2$D$^+$, as 
opposed to colder regions ($\lesssim 20$~K) like the Horsehead dense-core, 
where H$_2$D$^+$ is the main actor. Owing to the low temperature in the core 
it is likely that a non-negligible fraction of CO is frozen on the dust 
grains, enhancing the deuterium fractionation. 

Another way to form deuterated molecules in cold environments is
trough D addition or H-D substitution reactions on the surface of dust
grains \citep{hidaka09}. In the Horsehead core though, desorption from
the grain mantles is not efficient in releasing products
into the gas-phase (see section~\ref{sec:chemistry}). It is then more
likely that the gas-phase HDCO and $\ddco$ molecules detected
here are formed in the gas-phase. Nevertheless, there can still be a
considerable amount of deuterated $\hhco$ trapped in the ices around dust
grains.

\section{H$_2$CO chemistry}
\label{sec:chemistry}

We used a one-dimensional, steady-state photochemical model 
\citep{lebourlot93,lepetit06} to study the $\hhco$ chemistry in the Horsehead. 
The physical conditions have already been constrained by our previous
observational studies and we keep the same assumptions for the density
profile (displayed in the upper panel of Fig.~\ref{fig:meudon}), radiation
field ($\chi = 60$ in Draine units), elemental gas-phase abundances
\citep[see Table 6 in][]{goicoechea09b}
and cosmic ray ionization rate ($\zeta = 5 \times 10^{-17} \ps$).  

Unlike other organic molecules like methanol, which can only be efficiently 
formed on the surface of grains \citep{tielens97,woon02,cuppen09}, 
formaldehyde can be formed in both the gas-phase and on the surface of grains.
Next, we investigate these two different scenarios.

\subsection{Pure gas-phase chemistry models}

\FigMeudon{}%

We use dthe \textit{Ohio State University (osu)} pure gas-phase chemical 
network upgraded to photochemical studies. We included the 
photo-dissociation of HCO and of $\hhco$ (leading to CO and $\hh$) with 
rates of  $1.1\times10^{-9} \exp(-0.8 A_V)$ and $10^{-9} \exp(-1.74 A_V) 
\ps$, respectively \citep{vandishoeck88}.
We also included the $\hhco$ photo-dissociation channel that leads to 
HCO and H \citep[see \eg,{}][]{yin07,troe07} with the same rate of the one 
that leads to CO and $\hh$, and the atomic oxygen reaction with the 
methylene radical (CH$_2$) to explain the high abundance of 
HCO in the PDR \citep{gerin09}.  

The predicted HCO and $\hhco$ abundance profiles and the HCO/$\hhco$ abundance 
ratio are shown as solid lines in Fig.~\ref{fig:meudon} (middle and lower 
panel, respectively). The formation of $\hhco$ in the PDR and dense-core 
is dominated by reactions between oxygen atoms and the methyl radical 
(CH$_3$). The destruction of $\hhco$ in the PDR is dominated by
photo-dissociation, while it is dominated by reactions with ions in
the dense-core. The pure-gas phase model satisfactorily reproduces the
observed $\hhco$ abundance in the dense-core ($\delta x \sim 35^{''}$)
but it predicts an abundance in the PDR ($\delta x \sim 15^{''}$) that
is at least one order of magnitude lower than the observed value.  

\subsection{Grain chemistry models}

We considered the surface chemistry reactions introduced by
\cite{stantcheva02}, which include the following sequence of hydrogen
addition reactions on CO to form formaldehyde and methanol 

\begin{center}
  CO $\xrightarrow[\h]{}$ HCO $\xrightarrow[\h]{} \hhco \xrightarrow[\h]{}
  \hhhco \xrightarrow[\h]{} \chhhoh$. 
\end{center}
We also introduce water formation via hydrogenation reactions of O, OH
until $\hho$.

Adsorption, desorption and diffusive reactions were introduced in
the Meudon PDR code in the rate equations approach. The corresponding
implementation will be described in a specific paper (Le Bourlot et
al., to be submitted) and we simply mention the main processes
included in the present study. We distinguish between mantle
molecules, which may accumulate in several layers (\eg, $\hho, \hhco,
\chhhoh$), and light species (\eg, H, $\hh$), which stay on the
external layer. Photo-desorption can be an efficient mechanism to
release molecules to the gas phase in regions exposed to strong
radiation fields, as shown recently in laboratory studies
\citep{oberg09b,oberg09a,munoz-caro10}. Thermal desorption is also
introduced. It critically depends on the desorption barrier
values, which are somewhat uncertain. Diffusive reactions occur on
grain surfaces and the diffusion barriers are assumed to be 1/3 of the
desorption energy values. Photodesorption efficiencies have been 
measured in the laboratory for CO, CO$_2$, H$_2$O and $\chhhoh$. These
experiments have shown that all common ices have photodesorption
yields of a few $10^{-3}$ molecules per incident UV photon
\citep{oberg07,oberg09a,oberg09b,oberg09c}. Therefore, we also take
a photo-desorption efficiency of $10^{-3}$ for 
those species that have not been studied in the laboratory. We assume
in addition that for formaldehyde the two branching ratios toward
$\hhco$ and HCO+H channels are identical, \ie{}
$5\times10^{-4}$. Given the high density in the dense-core, the grains
are assumed to be strongly coupled to the gas in the inner region, so
that their temperatures become equal to 20 K in the dark region,
whereas the illuminated dust grains reach temperature values of about
30 K.       

The predicted HCO and $\hhco$ abundances are shown as dashed lines in 
Fig.~\ref{fig:meudon}. This model reproduces the observed $\hhco$ abundance 
in the dense-core and predicts a similar abundance as the pure
gas-phase model. This way, formation on grain surfaces does not
contribute significantly to the observed gas-phase $\hhco$ abundance
in the dense-core. This is because of the low photo-desorption
rates in the core caused by the shielding from the external 
UV field. On the other hand, the $\hhco$ abundance can increase by
up to three orders of magnitude in the illuminated part of the cloud ($A_V
\lesssim 4$) when including the grain surface reactions. The $\hhco$
abundance now even peaks in the PDR, while it peaked in the dense-core
in the pure gas-phase model. The model predicts a $\hhco$ abundance
peak in the PDR that is higher than the observed abundance averaged
over the 30m ($\sim16"$). This limited resolution prevents us from
resolving the predicted abundance peak. Interferometric observations
are needed to prove the existence of this peak in the PDR.

\section{Discussion}
\label{sec:discussion}

$\hhco$ has been detected in a variety of different astrophysical
environments, with a wide range of gas temperatures and densities. It
has been detected in diffuse clouds with high abundances ($\sim
10^{-9}$), observed in absorption against bright HII regions
\citep[\eg,][]{liszt95,liszt06}. It is not well understood how 
$\hhco$ can be formed and survive in such harsh environments, because
gas-phase process cannot compete with the photo-dissociation and dust
grain temperatures are too high for molecules to freeze on their
surfaces. \cite{roueff06} detected absorption lines of $\hhco$ at
3.6$\mu$m toward the high-mass protostar W33A, and estimated an
$\hhco$ abundance of $\sim10^{-7}$ where the gas has a temperature of 
$\sim100$~K. Recently, \cite{bergman11} found $\hhco$ abundances 
$\sim5\times10^{-9}$ in the $\rho$ Ophiuchi A cloud core.
Abundances of $\hhco$ and other more complex molecules toward hot cores 
and protostars are high. In these regions the gas is dense and hot,
so the dust grains also have high temperatures ($>100$ K). Therefore,
the ice mantles, formed in the cold pre-stellar phase, are completely
evaporated. Once these molecules are in the gas-phase, they trigger an
active chemistry in the hot gas, forming even more complex molecules
\citep{charnley92}.
 
$\hhco$ has also been observed in other PDRs. \cite{leurini10}
detected $\hhco$ in the Orion Bar PDR toward both the clump ($\nH
\sim 10^6 \pccm$) and the inter-clump ($\nH \sim 10^4 \pccm$) gas
components. They found higher $\hhco$ abundances ($\sim 10^{-9} -
10^{-7}$) than the ones inferred in this work for the Horsehead
($\sim10^{-10}$). Molecules trapped in the ice mantles can be
thermally desorbed when the dust grains are warm enough. The dust 
temperature at which a significant amount of $\hhco$ evaporates can
be estimated by equating the flux of desorbing molecules from the
ices to the flux of adsorbing molecules from the gas \citep[see
eq. 5 in][]{hollenbach09}. Taking an $\hhco$ desorption energy of
2050~K \citep{Garrod06}, we obtain an evaporation temperature of
$\sim 41$~K. In the Orion Bar the dust grains have temperatures of 
$\Td>55-70$~K, so molecules can be desorbed from the icy mantles
both thermally and non-thermally. But in the Horsehead PDR 
dust grains are colder ($\Td \sim 20-30$~K), therefore molecules can only
be desorbed non-thermally. Hence, the main desorption mechanism in
the PDR is photo-desorption. In this respect, the Horsehead PDR
offers a cleaner environment to isolate the role of FUV
photo-desorption of ice mantles. In the Horsehead dense-core dust
grains are also cold ($\sim20$ K), but photo-desorption is not
efficient because the dust is shielded from the external UV
field. Cosmic rays can desorb molecules from the ice mantles, but
this contribution is not significant because the desorption rates
are too low compared to the $\hhco$ formation rates in the gas-phase.  
Both the measured $\hhco$ abundance and ortho-to-para ratio 
agree with the scenario in which $\hhco$ in the dense-core 
is formed in the gas phase with no significant contribution
from grain surface chemistry.

We have shown that photo-desorption is an efficient mechanism to form 
gas-phase $\hhco$ in the Horsehead PDR. But, to understand
the importance of grain surface chemistry over gas-phase chemistry in
the formation of complex organic molecules, a similar analysis of
other molecules, such as $\chhhoh$ and C$\hhco$, is needed. In
particular, $\chhhoh$ is one of final products in the CO 
hydrogenation pathway on grain surfaces. It can also form $\hhco$
when it is photo-dissociated. Therefore, their gas-phase abundance
ratios will help us to constrain their dominant formation mechanism
and the relative contributions of gas-phase and grain surface
chemistry. Similar studies in different environments will also
bring additional information about the relative efficiencies of the
different desorption mechanisms. 

\section{Summary and conclusions}
\label{sec:conclusions}

We have presented deep observations of $\hhco$ lines toward the
Horsehead PDR  and a shielded condensation less than 40" away from the
PDR edge. We complemented these observations with a $\phhco$ emission
map. $\hhco$ emission is extended throughout the Horsehead with a
relatively constant intensity and resembles the 1.2~mm dust
continuum emission. $\hhco$ beam-averaged abundances are similar
($\simeq 2-3 \times 10^{-10}$) in the PDR and dense-core positions. 
We infer an equilibrium $\hhco$ ortho-to-para ratio of $\sim 3$ 
in the dense-core, while in the PDR we find a non-equilibrium value
of $\sim 2$.

For the first time we investigated the role of grain surface
chemistry in our PDR models of the Horsehead. Pure gas-phase and grain
surface chemistry models give similar results of the $\hhco$ abundance
in the dense-core, both consistent with the observations. This 
way, the observed gas-phase $\hhco$ in the core is formed mainly
trough gas-phase reactions, with no significant contribution from
surface process. In contrast, photo-desorption of $\hhco$
ices from dust grains is needed to explain the observed $\hhco$
gas-phase abundance in the PDR, because gas-phase chemistry alone does not 
produce enough $\hhco$. These different formation routes are
consistent with the inferred $\hhco$ ortho-to-para ratios.
Thus, photo-desorption is an efficient mechanism to produce
complex organic molecules in the PDR. Because the chemistries of
$\hhco$ and $\chhhoh$ are closely linked, we will continue this
investigation in a next paper by studying the chemistry of $\chhhoh$
in detail.    

\begin{acknowledgements}
  We thank A. Faure and N. Troscompt for sending us the $\phhco$ - 
  $\ohh$ and $\phhco$ - $\phh$ collisional rates prior to
  publication. We thank the referee for a careful reading of the
  manuscript and interesting comments. VG thanks support from the
  Chilean Government through the Becas Chile scholarship program. This
  work was also funded by grant ANR-09-BLAN-0231-01 from the French
  {\it Agence Nationale de la Recherche} as part of the SCHISM
  project. JRG thanks the Spanish MICINN for funding support through
  grants AYA2009-07304 and CSD2009-00038. JRG is supported by a
  Ram\'on y Cajal research contract from the Spanish MICINN and
  co-financed by the European Social Fund. 
\end{acknowledgements}

\bibliographystyle{aa}
\bibliography{draft_hh_h2co}

\begin{appendix}

  \section{$\hh$ ortho-to-para ratio}
  \label{app:1}

\begin{figure}[h!]
    \centering
    \includegraphics[scale=.6]{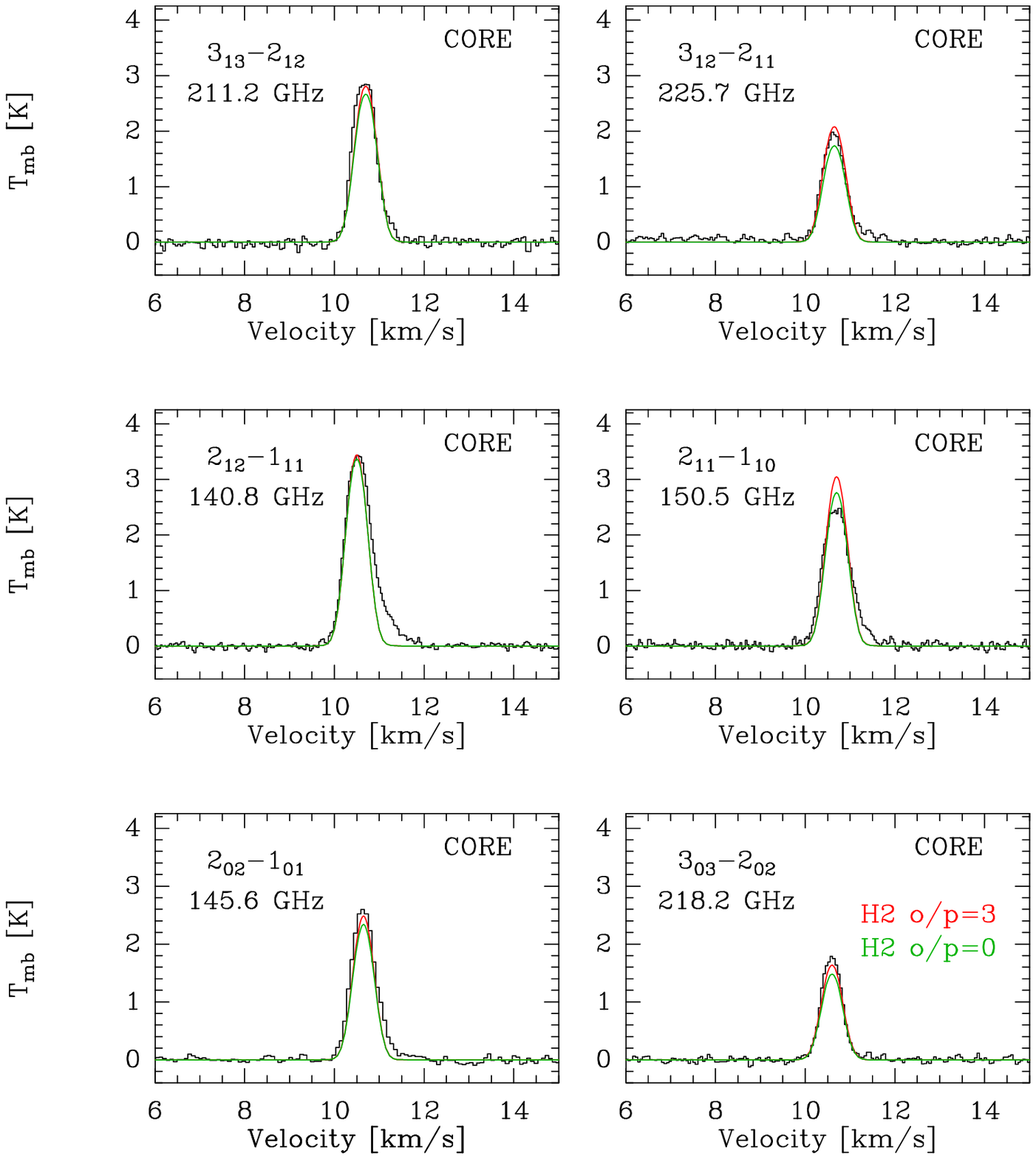}
    \caption[Optional caption for list of figures]{Radiative-transfer modeling 
      of $\hhco$ lines for the core position in the Horsehead. The two
      top rows display the ortho lines and the bottom row displays the
      para lines. The best-match models are given in colors ($\Tkin=20$ K,
      $n(\hh)~=~10^5 \pccm$, N($\ohhco$)~=~$9.6~\times~10^{12} \pscm$,
      N($\phhco$)~=~$3.2~\times~10^{12} \pscm$), taking a $\hh$
      ortho-to-para ratio of 3 (red lines) and of 0 (green lines).}
      \label{fig:h2opratio}
  \end{figure}

  We investigated the influence of the $\hh$ ortho-to-para ratio
  adopted in the excitation and radiative transfer models. In
  Fig.~\ref{fig:h2opratio} we show the best-match models for the
  $\hhco$ lines toward the core position in the Horsehead assuming
  two different values for the $\hh$ ortho-to-para ratio. We show
  models for an $\hh$ ortho-to-para ratio of 3 in red (high
  temperature limit), and we show models for the extreme case where
  the $\hh$ ortho-to-para ratio is 0 in green (low temperature
  limit). The difference between the models is less than 10$\%$, 
  which is within the observational uncertainties and therefore  not 
  significant.

\end{appendix}


\end{document}